\documentclass[12pt,draftcls,onecolumn]{IEEEtran}
\ifCLASSINFOpdf
\usepackage[pdftex]{graphicx}
\graphicspath{{../pdf/}{../jpeg/}}
\DeclareGraphicsExtensions{.pdf,.jpeg,.png}
\else
\usepackage[dvips]{graphicx}
\graphicspath{{../eps/}}
\DeclareGraphicsExtensions{.eps}
\fi\usepackage{graphicx}

\usepackage{graphics}
\usepackage{epsfig}
\usepackage{epstopdf}
\usepackage{stfloats}
\usepackage[cmex10]{amsmath}
\usepackage{algorithmic}
\usepackage{array}
\usepackage{mdwmath}
\usepackage{mdwtab}
\usepackage{graphicx}
\usepackage{subfigure}
\usepackage{color}
\usepackage{amsfonts,amssymb}
\usepackage{hyperref} 
\usepackage{multirow}
\usepackage{diagbox}
\usepackage{caption}

\usepackage{amsfonts,amsthm,array} 
\usepackage[ruled]{algorithm2e}

\begin{document}
\title{Joint Trajectory Design and User Scheduling of Aerial Cognitive Radio Networks\thanks{Manuscript received.}}

\author{Hongjiang~Lei, 
	Haosi~Yang,
	Ki-Hong~Park, 
	Imran~Shafique~Ansari, \\
	Jing~Jiang, 
	and~Mohamed-Slim~Alouini} 

\maketitle
	
\begin{abstract}
Unmanned aerial vehicles (UAVs) have been widely employed to enhance the end-to-end performance of wireless communications since the links between UAVs and terrestrial nodes are line-of-sight (LoS) with high probability. However, the broadcast characteristics of signal propagation in LoS links make it vulnerable to being wiretapped by malicious eavesdroppers, which poses a considerable challenge to the security of wireless communications. This paper investigates the security of aerial cognitive radio networks (CRNs). An airborne base station transmits confidential messages to secondary users utilizing the same spectrum as the primary network. An aerial base station transmits jamming signals to suppress the eavesdropper to enhance secrecy performance. The uncertainty of eavesdropping node locations is considered, and the average secrecy rate of the cognitive user is maximized by optimizing multiple users' scheduling, the UAVs' trajectory, and transmit power. To solve the non-convex optimization problem with mixed multiple integers variable problem, we propose an iterative algorithm based on block coordinate descent and successive convex approximation. Numerical results verify the effectiveness of our proposed algorithm and demonstrate that our scheme is beneficial to improving the secrecy performance of aerial CRNs.
		
\end{abstract}
	

\section{Introduction}
\label{sec:introduction}

\subsection{Background and Related Works}
\label{sec:Background and Related Works}

Because of high maneuverability, low cost,  and flexibility for on-demand deployment, low-altitude unmanned aerial vehicles (UAVs) are extensively utilized in diverse fields for different applications and purposes, such as real-time surveillance, traffic control, and communication relays because of high maneuverability, low cost, and flexibility for on-demand deployment, policing, and inspection, etc.
Among all these application scenarios, UAV communication is an up-and-coming technology that will become an essential part of future mobile communication systems \cite{WuQ2021JSAC, LiB2021IoT}.
Expressly, by adjusting the position or designing the trajectory of UAVs, a reliable line-of-sight link (LoS) can be provided to the terrestrial nodes (TNs) with high possibility.
Therefore, the altitude and the horizontal location of UAVs have a decisive effect on the performance of UAV communication systems.
Trajectory design has become a critical issue to resolve in UAV communication system design \cite{WuQ2019WC22}.

The most typical scenario of UAV-assisted wireless communications is providing ubiquitous wireless coverage within the serving area for a given geographical area \cite{ZengY2016Mag}, \cite{ZengY2019Proc}.
A communication system with multiple aerial base stations (BSs) was investigated in \cite{WuQ2018TWC} in which UAV trajectories were optimized to maximize the minimum throughput concerning user scheduling coefficients and transmitting power on UAVs.
The authors proposed an effective iterative algorithm based on the successive convex approximation (SCA) and the block coordinate descent (BCD) to solve the mixed-integer non-convex optimization problem.
The results demonstrated that UAV communication systems could enhance the system utility with additional flexibility for interference mitigation.
Considering a finite amount of energy and a certain quality of service requirement, the max-min fairness problem was studied in \cite{BejaouiA2020WCL}.
When UAVs are utilized to be relay nodes, long-distance wireless links between terrestrial BSs and TNs can be quickly established, which is of great use in disaster or other emergency scenarios \cite{ZengY2016Mag}.
In \cite{LiuT2021TGCN}, the authors considered a UAV-assisted cooperative system in which multiple UAVs were utilized as decode-and-forward relays to forward signals to the TNs.
The minimum transmission rate of the system was maximized by jointly optimizing the flight trajectory of the relay UAVs, the transmitting power on UAVs, and terrestrial BS.
An iterative algorithm was proposed to solve the joint optimization problem based on the block coordinate ascent technique, introducing the slack variable and SCA techniques.
In \cite{LeeJH2021TCOM}, an aerial relay balanced the difference in transmission rates between free space optical and radio frequency links, and the throughput for delay-limited and delay-tolerant scenarios was maximized by carefully designing the flight trajectory of the aerial relay, which in turn maximized the throughput.
In UAV-aided wireless sensor networks and the IoT communication systems, UAVs are utilized as aerial access points to disseminate/collect information to/from TNs \cite{ZengY2016Mag}, \cite{ZengY2019Proc}.
A mixed-integer non-convex optimization problem was proposed to prolong the wireless sensor network lifetime and the UAV's trajectory were jointly optimized to minimize the maximum energy consumption of all sensor nodes in \cite{ZhanC2018WCL}. An efficient iterative algorithm was proposed to obtain a sub-optimal solution by applying the SCA technique.
In \cite{LuW2021CIS}, two UAVs were utilized to provide energy for two TNs and collect information from the two TNs alternately. A resource and trajectory optimization scheme was proposed to maximize the minimum throughput of TNs by joint optimization of UAVs' trajectories, time allocation, and TNs' transmit power.
The minimum data collection rate was maximized by jointly optimizing the 3D trajectory of UAVs and time allocation in \cite{LuoW2021IoT}. The minimum distances between UAVs were considered to avoid a collision and an efficient iterative algorithm was proposed to solve the non-convex optimization problem.

Information security is one of the fundamental requirements for wireless communication systems.
High possibilities of LoS air-to-ground (A2G) communication links make UAV-assisted wireless communication systems more vulnerable and more prone to be eavesdropped by malicious nodes \cite{WuQ2019WC, WangHM2019WC, LiB2019WC}.
In \cite{LiZ2019CL},  the secrecy performance of a UAV-aided wireless communication system with multiple TNs was investigated and the minimum secrecy rate was maximized by jointly optimizing aerial BS's trajectory, the transmit power, and the users association.
An iterative algorithm was proposed by the SCA technique and the alternating method, and their results showed that the secrecy rate was improved.
Because of LoS links with a weak path loss and high maneuverability, UAVs can be utilized as aerial jammers to suppress the eavesdroppers \cite{LeeH2018TVT, ZhongC2019CL, ZhangR2021TWC}.
In \cite{LeeH2018TVT}, the physical layer security issues of a dual-UAV communication system with multiple TNs were investigated, and the minimum secrecy rate was maximized by jointly optimizing the trajectories and the transmit powers of both UAVs as well as the user scheduling.
However, the location of the eavesdropper is assumed to be available to the base station in \cite{LiZ2019CL} and \cite{LeeH2018TVT}.
In \cite{ZhongC2019CL}, the secrecy performance of a wireless system with one aerial BS and one cooperative aerial friendly jammer was investigated, and the average secrecy rate was maximized by jointly optimizing UAVs' trajectory and the transmit power. The location of the TN was assumed to be perfectly known while the location of the eavesdropper was assumed to be partially known for both UAVs.
An iterative algorithm was proposed to solve the optimization problem using the alternating optimization and SCA techniques.
In comparison, Zhang \emph{et al.} investigated a dual-UAV communication system with multiple TNs in \cite{ZhangR2021TWC}. Both the average secrecy rate and the secrecy energy efficiency were maximized by jointly optimizing the user scheduling, the transmit power, the trajectory, and the velocity of the two UAVs.

Although UAV-enabled cognitive radio networks (CRNs) can improve spectrum efficiency compared to traditional wireless communication networks, how to effectively mitigate the interference from A2G co-channel to the primary links is an important and challenging problem \cite{HuangY2019TCOM, UllahZ2020TCCN}.
A UAV-enabled CRN with one cognitive user (CU), multiple eavesdroppers, and multiple primary users (PUs) was considered in \cite{ZhouYF2020TCOM} and the average secrecy rate of the CU in the worst-case scenario was maximized by optimizing the aerial BS's trajectory and transmit power.
The location of PUs and eavesdroppers is assumed to be inaccurate and iterative algorithms based on bounded and probabilistic location error models were proposed to obtain a suboptimal solution.
Wang \emph{et al.} considered the secrecy performance of a CRN with multiple terrestrial CUs and PUs, and a terrestrial eavesdropper in \cite{WangY2021TCCN}. An aerial-friendly jammer was utilized to enhance security.
The total average secrecy rate of the CU was maximized by jointly optimizing the sub-carrier allocation, the trajectory, and the transmission power of the aerial jammer and the power of the cognitive BS.
In \cite{NguyenPX2021TVT}, the authors investigated the secrecy performance of a CRN with a terrestrial CU, a PU, an eavesdropper, and a friendly jammer. Two scenarios were considered in which all the location of ground nodes (including the CU, PU, and eavesdropper) was assumed to be known and unknown, respectively. The average achievable secrecy rate of the CU was maximized by jointly optimizing the transmit power and three-dimensional (3D) trajectory of the aerial jammer.

\subsection{Motivation and Contributions}
The combination of physical layer security technology and cognitive wireless networks on UAV platforms exploits the advantages of high mobility, security, flexibility, low cost, and on-demand distribution of UAVs. It efficiently utilizes limited spectrum resources, guaranteeing high-quality transmission during wireless communication.
Motivated by this practical significance, in this work, we investigate the dual collaborative UAV security scheduling for multi-user transmission under the uncertainty of potential eavesdropping node locations in a CRN. The main contributions of this paper are summarized as follows:

\begin{enumerate}
	\item We propose a dual collaborative UAV system for physical layer security in a multi-user scheduling scenario under the uncertainty of potential eavesdropping node locations in CRNs. A legitimate UAV operates as an airborne base station and transmits confidential messages to secondary users utilizing the same spectrum licensed to the primary network. A friendly jamming UAV disrupts eavesdropping by transmitting artificial noise. Meanwhile, a malicious eavesdropping node at an uncertain location tries to wiretap the confidential information. In this dual-UAV network, the average secrecy rate is maximized by jointly optimizing the user scheduling, the transmit power of UAVs, and the trajectory of the UAVs.
	
	\item The average sum secrecy rate is formulated by jointly optimizing the scheduling, the transmit power, and the trajectory of the two UAVs subject to the limited power constraints,  the maximum flying speed, initial and final position restrictions of UAVs, and threshold for the regular operation of the primary network. Due to the non-convexity, the challenging optimization problem is divided into several subproblems while transformed into approximated convex forms via the SCA. Subsequently, the BCD technology is employed to address these subproblems successively.
	
	\item  We compare the proposed algorithm with other benchmark strategies, which only consider the fixed trajectory and transmit power of UAVs while considering only the case of optimizing individual UAV trajectory and transmit power. The simulation results verify the efficiency and the convergence of the proposed algorithm.
\end{enumerate}

\subsection{Organization}
The rest of this paper is organized as follows.
The system model and problem formulation are provided in Section \ref{sec:SystemModel}.
Section \ref{sec:ProposedAlgorithm1} presents a proposed iterative algorithm for the case of friendly jammer drones to solve the same.
Simulation results are demonstrated in Section \ref{sec:Simulation}. Finally, Section \ref{sec:Conclusions} concludes this paper.

\section{System Model and Problem Formulation}
\label{sec:SystemModel}
\subsection{System Model}

\begin{figure}[t]
	\centering		
	\includegraphics[width = 3.5 in]{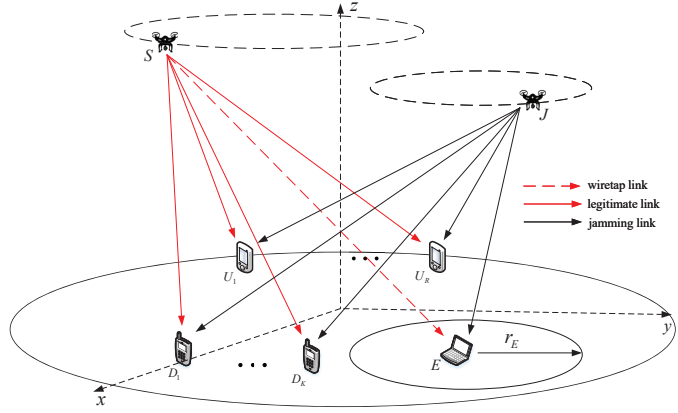}
    \caption{Dual-UAV enabled secure communication system in CRN.}
    \label{fig_model}
\end{figure}

As shown in Fig. \ref{fig_model}, we consider an underlay CRN with an airborne base station (${S}$) transmitting confidential messages to CUs utilizing the same spectrum licensed to the primary network.
A malicious terrestrial eavesdropper ($E$) at uncertain location is trying to wiretap the confidential information.
A  jammer UAV (${J}$) sends jamming signals to suppress the eavesdropping capabilities of $E$.
It is assumed that there are $R$ PUs (${U_r}, r = 1, \cdots, R$) and $K$ CUs (${D_k}, k = 1, \cdots, K$) where all devices are equipped with a single antenna.
Similar to \cite{ZhongC2019CL}, ${S}$ and ${J}$ have only partial information about $E$ due to the uncertainty of its location.
A three-dimensional Cartesian coordinate is utilized where ${S}$ and ${J}$ fly at a fixed altitude ${H_1}$ and ${H_2}$, respectively.
The flight period of two UAVs, ${T}$, is divided into ${N}$ time slots as ${\delta _t} = \frac{T}{N}$.
When ${\delta _t}$ is small enough, the position of  UAVs at each point can be approximated as a constant \cite{WuQ2018TWC}.
The horizontal positions of ${S}$ and ${J}$ at the $n$th slot are expressed as ${{\mathbf{q}}_S}\left( n \right) = {\left[ {{x_S}\left( n \right),{y_S}\left( n \right)} \right]^T}$ and ${{\mathbf{q}}_J}\left( n \right) = {\left[ {{x_J}\left( n \right),{y_J}\left( n \right)} \right]^T}$, respectively.
In addition, ${{\bf{w}}_{{D_k}}} \in {\mathbb{R}^{2 \times 1}}$, ${{\bf{w}}_{{U_r}}} \in {\mathbb{R}^{2 \times 1}}$, and ${{\mathbf{w}}_E} \in {\mathbb{R}^{2 \times 1}}$ denote the location coordinates of ${D_k}$, ${U_r}$, and $E$.
Similar to \cite{ZhangR2021TWC} and \cite{WangY2021TCCN}, all the A2G links are assumed to be LoS links.
For simplicity, it is assumed that the A2G links are dominated by the LoS links, where the channel quality depends only on Euclidean distance. Furthermore, the Doppler effect is also assumed to be well compensated by the receivers.
The channel coefficients between $S$ and all the receivers ($D_k$, $U_r$, and $E$) at the $n$th slot are expressed as
\begin{subequations}
	\begin{align}
        {h_{S{D_k}}}\left( n \right) &= \frac{{{\rho _0}}}{{{{\left\| {{{\mathbf{q}}_S}\left( n \right) - {{\bf{w}}_{{D_k}}}} \right\|}^2} + H_1^2}},  \label{hSk}    \\
        {h_{S{U_r}}}\left( n \right) &= \frac{{{\rho _0}}}{{{{\left\| {{{\mathbf{q}}_S}\left( n \right) - {{\bf{w}}_{{U_r}}}} \right\|}^2} + H_1^2}},  \label{bSr}	\\
        {h_{SE}}\left( n \right) &= \frac{{{\rho _0}}}{{{{\left\| {{{\mathbf{q}}_S}\left( n \right) - {{\mathbf{w}}_E}} \right\|}^2} + H_1^2}},   \label{hSE}
	\end{align}
\end{subequations}
where $\rho _0$ denotes the channel power gain at the reference distance.
Similarly, the channel coefficients between $J$ and all the receivers at the $n$th slot are expressed as
\begin{subequations}
	\begin{align}
        {h_{J{D_k}}}\left( n \right) &= \frac{{{\rho _0}}}{{{{\left\| {{{\mathbf{q}}_J}\left( n \right) - {{\bf{w}}_{{D_k}}}} \right\|}^2} + H_2^2}}, \label{hJk}	\\
        {h_{J{U_r}}}\left( n \right) &= \frac{{{\rho _0}}}{{{{\left\| {{{\mathbf{q}}_J}\left( n \right) - {{\bf{w}}_{{U_r}}}} \right\|}^2} + H_2^2}}, \label{hJr} 	\\
        {h_{{JE}}}\left( n \right) &= \frac{{{\rho _0}}}{{{{\left\| {{{\mathbf{q}}_J}\left( n \right) - {{\mathbf{w}}_E}} \right\|}^2} + H_2^2}}.   \label{hJE}
	\end{align}
\end{subequations}
Since the exact location of $E$ is not available for ${S}$ and ${J}$, ${{h_{SE}}}$ and ${{h_{JE}}}$ in (\ref{hSE}) and (\ref{hJE}) are inaccurate, ${{\mathbf{\hat w}}_E} \in {\mathbb{R}^{2 \times 1}}$ and ${r_E}$ are utilized to indicate the circle center and radius of the area wherein $E$ is located. 
It is assumed that ${r_E}$ is smaller than the distance between $S$ and $E$, i.e. $\left\| {{{\bf{q}}_S}\left( n \right) - {{{\bf{\hat w}}}_E}} \right\| \ge {r_E}$. 
According to the triangle inequality, the distance between ${S}$ and $E$ is expressed as
\begin{equation}
 	\begin{aligned}
		\left\| {{{\mathbf{q}}_S}\left( n \right) - {{\mathbf{w}}_E}} \right\| &\geqslant \left| {\left\| {{{\mathbf{q}}_S}\left( n \right) - {{{\mathbf{\hat w}}}_E}} \right\| - \left\| {{{{\mathbf{\hat w}}}_E} - {{\mathbf{w}}_E}} \right\|} \right| \\
		& \geqslant \left| {\left\| {{{\mathbf{q}}_S}\left( n \right) - {{{\mathbf{\hat w}}}_E}} \right\| - {r_E}} \right|.
		\label{DBE}	
	\end{aligned}	
\end{equation}
When ${r_E} = 0$, it indicates that the base station is fully able to obtain the location information of $E$.
At this point, the whole problem degenerates into the physical layer security problem of the UAV trajectory design based on the perfect eavesdropping position considered in \cite{LiZ2019CL}.

Similarly, the distance between ${J}$ and $E$ is expressed by the upper and lower bounds, respectively, as follows.
\begin{equation}
	\begin{aligned}
		\left\| {{{\mathbf{q}}_J}\left( n \right) - {{\mathbf{w}}_E}} \right\| &\leqslant \left\| {{{\mathbf{q}}_J}\left( n \right) - {{{\mathbf{\hat w}}}_E}} \right\| + \left\| {{{{\mathbf{\hat w}}}_E} - {{\mathbf{w}}_E}} \right\| \hfill \\
		& \leqslant \left\| {{{\mathbf{q}}_J}\left( n \right) - {{{\mathbf{\hat w}}}_E}} \right\| + {r_E}.
		\label{DJE}
	\end{aligned}
\end{equation}
Then ${{h_{SE}}}$ and ${{h_{JE}}}$  is rewritten as
\begin{equation}
	{h_{SE}}\left( n \right) = \frac{{{\rho _0}}}{{{{\left( {\left\| {{{\mathbf{q}}_S}\left( n \right) - {{{\mathbf{\hat w}}}_E}} \right\| - {r_E}} \right)}^2} + H_1^2}}
\end{equation}
and
\begin{equation}
	{h_{{JE}}}\left( n \right) = \frac{{{\rho _0}}}{{{{\left( {\left\| {{{\mathbf{q}}_J}\left( n \right) - {{{\mathbf{\hat w}}}_E}} \right\| + {r_E}} \right)}^2} + H_2^2}},
\end{equation}
respectively.

To enable all the CUs to be served fairly in one period, a binary variable ${\theta _k}\left( n \right)$ is utilized to characterize user scheduling in the considered CRN \cite{ChenY2019AS}.
More specifically, ${\theta _k}\left( n \right)= 1 $ signifies that only $D_k$ is scheduled to communicate with ${S}$ at the ${n}$th slot.
Thus, we have	
\begin{equation}
	\sum\limits_{k = 1}^K {{\theta _k}\left( n \right)}  \le 1,\forall n,
\end{equation}	
\begin{equation}
	{\theta _k}\left( n \right) \in \left\{ {0,\left. 1 \right\}} \right.,\forall n,k.
\end{equation}	

It is assumed that ${S}$ for the $k$th user and ${J}$ transmit signals at the ${n}$th slot with power ${P_k}\left( n \right)$ and ${{P_J}\left( n \right)}$, respectively. The average interference power constraint for $U_{r}$ is given as {\cite{WangY2021TCCN}}
\begin{equation}
\frac{1}{N}\sum\limits_{n = 1}^N {\left\{ {{P_J}\left( n \right){h_J}_{{U_r}}\left( n \right) + \sum\limits_{k = 1}^K {{\theta _k}\left( n \right){P_k}\left( n \right){h_S}_{{U_r}}\left( n \right)} } \right\}}  \leqslant {\Gamma _r},
\end{equation}	
where ${\Gamma _r}$ denotes a constant threshold value of the ${U_r}$ that can tolerate the maximum average interference power.

In this work, it is assumed that
the jamming signals transmitted by $J$ can be a Gaussian pseudo-random sequence or the same utilizes deterministic waveforms similar to the structure of the desired signal as \cite{LvL2019TIFS}, \cite{CaiY2018JSAC, XingH2016TVT}. Then
the jamming signals sent by ${J}$ can be cancelled by the received signal at all the secondary receivers, the signal-to-noise ratio (SNR) for ${D_k}$ is expressed as
\begin{equation}
	{\gamma _{D_k}} = \frac{{{P_k}\left( n \right){h_{S{D_k}}}}}{{{\sigma ^2}}},
\end{equation}
where ${{\sigma ^2}}$ denotes the variance of additive white Gaussian noise (AWGN).
The achievable rate ${R_{S{D_k}}}\left( n \right)$ is expressed as
\begin{equation}
	{R_{S{D_k}}}\left( n \right) = {\log _2}\left( {1 + \frac{{{P_k}\left( n \right){h_{S{D_k}}}}}{{{\sigma ^2}}}} \right).
\end{equation}
Similarly, the signal-to-interference-plus-noise ratio (SINR) and achievable rate at $E$ are expressed as
\begin{equation}
	{\gamma _E} = \frac{{{P_k}\left( n \right){h_{SE}}}}{{{P_J}\left( n \right){h_{JE}} + {\sigma ^2}}}
\end{equation}
and
\begin{equation}
    {R_E}\left( n \right) = {\log _2}\left( {1 + \frac{{{P_k}\left( n \right){h_{SE}}}}{{{P_J}\left( n \right){h_{JE}} + {\sigma ^2}}}} \right),
\end{equation}
respectively.
Then the secrecy rate at ${D_k}$ is expressed as
\begin{equation}
	{R_{\sec }^k}\left( n \right) = {\left[ {{{\log }_2}\left( {1 + \frac{{{P_k}\left( n \right){h_{S{D_k}}}}}{{{\sigma ^2}}}} \right) - {{\log }_2}\left( {1 + \frac{{{P_k}\left( n \right){h_{SE}}}}{{{P_J}\left( n \right){h_{JE}} + {\sigma ^2}}}} \right)} \right]^ + },
    \label{RSEC}
\end{equation}
where ${\left[ x \right]^ + } \triangleq \max \left( {x,0} \right)$.

\subsection{Problem Formulation}
In this work, the sum average secrecy rate is maximized with respect to UAV trajectory, the transmit power, and the user scheduling.
Let ${{\mathbf{\Theta }}} = \left\{ {{\theta _k}\left( n \right),\forall k,n} \right\}$, ${\mathbf{P}} = \left\{ {{P_k}\left( n \right),{P_J}\left( n \right),\forall k,n} \right\}$, and ${\mathbf{Q}} = \left\{ {{{\mathbf{q}}_S}\left( n \right),{{\mathbf{q}}_J}\left( n \right),\forall n} \right\}$. Therefore, we formulate the following optimization problem.
\begin{subequations}
	\begin{align}
		\mathcal{P}_{1} \,:\, \max\limits_{{{\mathbf{\Theta }},{\mathbf{P}},{\mathbf{Q}}}}\; &\frac{1}{N}\sum\limits_{n = 1}^N {\sum\limits_{k = 1}^K {{\theta _k}\left( n \right){{R_{\sec }^k}}\left( n \right)} },      \label{eq:C0}\\
		{\mathrm{s.t.}}\; &\sum\limits_{k = 1}^K {{\theta _k}\left( n \right)}  \le 1,\forall n,     \label{eq:C1}\\
		&{\theta _k}\left( n \right) \in \left\{ {0,\left. 1 \right\}} \right.,\forall n,k, \label{eq:C2}\\
		&\frac{1}{N}\sum\limits_{n = 1}^N {{\theta _k}\left( n \right)} R_{\sec }^k\left( n \right) \ge {R_{\min }}, \label{eq:C10}\\
		&0 \le {P_k}\left( n \right) \le {P_S^{max} },\forall k,n,    \label{eq:C3}\\
		&0 \le {\frac{1}{N}\sum\limits_{n = 1}^N {\sum\limits_{k = 1}^K {{\theta _k}\left( n \right){P_k}\left( n \right) \le } } P_S^{ave}},\forall k,n,   \label{eq:C4}\\
		&0 \le {P_J}\left( n \right) \le {P_J^{max} }, {\rm{for}}\;\forall n,   \label{eq:C5}\\
		&{\frac{1}{N}\sum\limits_{n = 1}^N {{P_J}\left( n \right)}  \le {P_J^{ave}}},  \label{eq:C6}\\
		&{{\mathbf{q}}_i}\left( 1 \right) = {\mathbf{q}}_i^0,{{\mathbf{q}}_i}\left( N \right) = {\mathbf{q}}_i^{\textrm{F}},\forall i \in \left\{ {S,J} \right\}, \label{eq:C7}\\
		&\left\| {{{\mathbf{q}}_i}\left( {n + 1} \right) - {{\mathbf{q}}_i}\left( n \right)} \right\| \le {\delta _t}{V_{i,max }},\forall i \in \left\{ {S,J} \right\},n \in N, \label{eq:C8}\\
		&\frac{1}{N}\sum\limits_{n = 1}^N {\left\{ {{P_J}\left( n \right){h_J}_{{U_r}}\left( n \right) + \sum\limits_{k = 1}^K {{\theta _k}\left( n \right){P_k}\left( n \right){h_S}_{{U_r}}\left( n \right)} } \right\}}  \le {\Gamma _r},\forall r,   \label{eq:C9}
	\end{align}
\end{subequations}
where
$P_S^{max}$ and $P_J^{max}$ denote the peak transmit power of UAVs,  $P_S^{ave}$ represents the average power of $S$ scheduling ${D_k}$, and $P_J^{ave}$ denote the average transmission power of $J$, and
${\mathbf{q}}_i^0$ and ${\mathbf{q}}_i^{\textrm{F}}$ are the initial and final positions of the UAVs, respectively.
Eqs. (\ref{eq:C1}) and (\ref{eq:C2}) are the scheduling constrains of ground users, respectively,
(\ref{eq:C10}) is the constraint that all users are scheduled fairly, where ${R_{\min }}$ is the minimum average security rate,
(\ref{eq:C3}) - (\ref{eq:C6}) are the peak and average power constraints of transmitting and jamming signals in each time slot, respectively,
(\ref{eq:C7}) denotes the constraints on the initial and final positions of the UAVs,
(\ref{eq:C8}) depicts the maximum flight distance between adjacent time slots during the flight,
and
(\ref{eq:C9}) considers the underlay condition.


Several factors make the original problem $\mathcal{P}_{1}$ challenging to solve.
First, the  operator ${\left[ x \right]^ + }$ makes the objective function non-smooth at zero value, and the optimization variable ${{\mathbf{\Theta }}}$ is binary and thus (\ref{eq:C1}) and (\ref{eq:C2}) involve integer constraints. Second, even with fixed user scheduling, $\mathcal{P}_{1}$ is still non-convex concerning transmitting power variable ${\mathbf{P}} $ and UAV trajectory variable ${\mathbf{Q}} $. Therefore, $\mathcal{P}_{1}$ is a mixed-integer non-convex problem, which is challenging to optimally solve in general.

\section{Proposed Algorithm On Dual Collaborative UAVs}
\label{sec:ProposedAlgorithm1}

According to the method used in \cite{ZhongC2019CL} and \cite{WangY2021TCCN}, we can  remove the operator ${\left[ x \right]^ + }$. Meanwhile, to solve $\mathcal{P}_{1}$, alternating optimization method is utilized to optimize the scheduling variable ${{\mathbf{\Theta }}}$, transmit power allocation ${{\mathbf{P }}}$, and trajectory of UAV ${{\mathbf{Q }}}$ in an alternating manner, by considering the others to be given.

\subsection{Subproblem 1.1: User Scheduling Optimization}
In this subsection, the scheduling variable ${\mathbf{\Theta }}$ with given $\left\{ {{\mathbf{P}},{\mathbf{Q}}} \right\}$ is optimized firstly.
The binary variable ${\theta _k}\left( n \right)$ is relaxed into continuous variables between 0 and 1 to restrain the binary constraint (\ref{eq:C2}).
Then $\mathcal{P}_{1}$ is reduced as
\begin{subequations}
	\begin{align}
        \mathcal{P}_{1.1} \,:\, \max\limits_{{\mathbf{\Theta }}, {\eta _k}}\; & \sum\limits_{k = 1}^K {\eta _k}  \\
			{\mathrm{s.t.}}\; &\frac{1}{N}\sum\limits_{n = 1}^N {{\theta _k}\left( n \right){R_{\sec }^k}\left( n \right)}  \geqslant {\eta _k} , \\
			&{\eta _k} \ge {R_{\min }}, \\
			& 0 \leqslant {\theta _k}\left( n \right) \leqslant 1,  \\
			& (\textrm{\ref{eq:C1}}) , (\textrm{\ref{eq:C4}}) , (\textrm{\ref{eq:C9}}),
			\label{eq:P1.1}
	\end{align}
\end{subequations}

where
\begin{equation}
{{R_{\sec }^k}}\left( n \right) = {\log _2}\left( {1 + \frac{{{P_k}\left( n \right){h_{S{D_k}}}}}{{{\sigma ^2}}}} \right) - {\log _2}\left( {1 + \frac{{{P_k}\left( n \right){h_{SE}}}}{{{P_J}\left( n \right){h_{JE}} + {\sigma ^2}}}} \right).
\label{Rsec}
\end{equation}
$\mathcal{P}_{1.1}$ is a standard linear programming (LP), which can be solved by existing optimization tools such as CVX.

\subsection{Subproblem 1.2: Transmit Power Optimization }
For any given user scheduling as well as the trajectory of ${S}$ and ${J}$, the optimization problem of transmit power is expressed as
\begin{subequations}
	\begin{align}		
        \mathcal{P}_{1.2} \,:\, \max\limits_{{\mathbf{P}}, {\eta _k}}\; & \sum\limits_{k = 1}^K {\eta _k}   \hfill \\
		{\mathrm{s.t.}}\; &\frac{1}{N}\sum\limits_{n = 1}^N {{\theta _k}\left( n \right){{R_{\sec }^k}}\left( n \right)}  \geqslant {\eta _k} , \label{eq:p1.2b}\\
		&{\eta _k} \ge {R_{\min }}, \\
		& (\textrm{\ref{eq:C3}}) - (\textrm{\ref{eq:C6}}), (\textrm{\ref{eq:C9}}). \label{eq:P1.2c}	
	\end{align}
\end{subequations}
where ${{ R}_{\sec }^k}\left( n \right)$ is rewritten by
\begin{equation}
	\begin{aligned}
		{{R}_{\sec }^k}\left( n \right) &= {\log _2}\left( {1 + \frac{{{P_k}\left( n \right){h_{S{D_k}}}}}{{{\sigma ^2}}}} \right) - {\log _2}\left( {{P_J}\left( n \right){h_{JE}} + {\sigma ^2} + {P_k}\left( n \right){h_{SE}}} \right) \\
		&+ {\log _2}\left( {{P_J}\left( n \right){h_{JE}} + {\sigma ^2}} \right).
		\label{Rsec1}
	\end{aligned}
\end{equation}	
It must be noted that (\ref{eq:p1.2b}) is a non-convex constraint due to the second term in (\ref{Rsec1}) which is a convex function with respect to ${\mathbf{P}}$.
To tackle the non-convex constraint, the first-order Taylor expansion is utilized to transform ${{ R}_{\sec }^k}\left( n \right)$ into
\begin{equation}
    \begin{aligned}
    {R}_{\sec}^{k,\textrm{lb}}\left( n \right) &= {\log _2}\left( {1 + \frac{{{P_k}\left( n \right){h_{S{D_k}}}}}{{{\sigma ^2}}}} \right) + {\log _2}\left( {{P_J}\left( n \right){h_{JE}} + {\sigma ^2}} \right) \\
    &-{\log _2}\left( {P_J^m\left( n \right){h_{JE}} + {\sigma ^2} + P_k^m\left( n \right){h_{SE}}} \right)\\
	&- \frac{{{h_{JE}}\left( {{P_J}\left( n \right) - P_J^m\left( n \right)} \right) + {h_{SE}}\left( {{P_k}\left( n \right) - P_k^m\left( n \right)} \right)}}{{\ln (2)\left( {{P_J}^m\left( n \right){h_{JE}} + {\sigma ^2} + {P_k}^m\left( n \right){h_{SE}}}\right)}},
	\label{Rlb}
	\end{aligned}
\end{equation}
where $\left\{ {P_J^m\left( n \right),P_k^m\left( n \right),\forall n} \right\}$ is a given feasible point in the ${m}$th iteration and the subscript `${\textrm{lb}}$' denotes lower bound.
For a given feasible point $\left\{ {P_J^m\left( n \right),P_k^m\left( n \right),\forall n} \right\}$,
${R}_{\sec}^{k,\textrm{lb}}\left( n \right)$
is a convex function with respect to the optimization variables ${{P_k}\left( n \right)}$ and ${{P_J}\left( n \right)}$.

By replacing ${{{R}_{\sec }^k}\left( n \right)}$ with ${ R}_{\sec }^{k,\textrm{lb}}\left( n \right)$, $\mathcal{P}_{1.2}$ is approximated as a convex optimization problem that can thus be solved by standard convex optimization techniques such as CVX.

\subsection{Subproblem 1.3: Optimizing Trajectory of $S$ }
In this section, we aim at optimizing the trajectory of ${S}$ with given
$\left\{ {{\mathbf{\Theta }},{\mathbf{P}}} \right\}$. $\mathcal{P}_{1}$ is approximated as
\begin{subequations}
	\begin{align}		
		\mathcal{P}_{1.3} \,:\, \max\limits_{{{\mathbf{q}}_S}, {\eta _k}}\; & \sum\limits_{k = 1}^K {\eta _k}  \\
        {\mathrm{s.t.}}\; & \frac{1}{N}\sum\limits_{n = 1}^N {{\theta _k}\left( n \right)} {R_{\sec, 1}^k}\left( n \right) \ge {\eta _k} ,\\
        &{\eta _k} \ge {R_{\min }}, \\
		& (\textrm{\ref{eq:C7}}) - (\textrm{\ref{eq:C9}}),
		\label{eq:P1.3}
	\end{align}
\end{subequations}
where
${R_{\sec, 1}^k}\left( n \right) = {\log _2}\left( {1 + \frac{{c\left( n \right){\rho _0}}}{{{{\left\| {{{\mathbf{q}}_S}\left( n \right) - {{\bf{w}}_{{D_k}}}} \right\|}^2} + H_1^2}}} \right) - {\log _2}\left( {1 + \frac{{t\left( n \right){\rho _0}}}{{{{\left( {\left\| {{{\mathbf{q}}_S}\left( n \right) - {{{\mathbf{\hat w}}}_E}} \right\| - {r_E}} \right)}^2} + H_2^2}}} \right)$,
$c\left( n \right) = \frac{{{P_k}\left( n \right)}}{{{\sigma ^2}}}$, and  $t\left( n \right) = \frac{{{P_k}\left( n \right)}}{{{P_J}\left( n \right){h_{JE}} + {\sigma ^2}}}$.
It must be noted that $\mathcal{P}_{1.3}$ is a non-convex problem since the function in ${R_{\sec,1}}\left( n \right) $ is a non-concave function with respect to ${{{\mathbf{q}}_S}\left( n \right)}$ and (\ref{eq:C9}) is a non-convex constraint. By introducing the slack variables and applying the successive convex approximation (SCA), $\mathcal{P}_{1.3}$ is written as
\begin{subequations}
	\begin{align}		
			\mathcal{P}_{1.3} \,:\, \max\limits_{{{\mathbf{q}}_S},{\eta _k} ,{d_{SE}},{d_{S{U_r}}}}\; & \sum\limits_{k = 1}^K {\eta _k} \\
			{\mathrm{s.t.}}\;  &\frac{1}{N}\sum\limits_{n = 1}^N {{\theta _k}\left( n \right)} \left( {{{\hat R}_{S{D_k}}}\left( n \right) - {{\log }_2}\left( {1 + \frac{{t\left( n \right){\rho _0}}}{{{d_{SE}} + H_2^2}}} \right)} \right) \ge {\eta _k}, \label{eq:p1.3b}\\
			&{\eta _k} \ge {R_{\min }}, \label{eq:p1.3g}\\
			&\frac{1}{N}\sum\limits_{n = 1}^N {\left\{ {{P_J}\left( n \right){h_{J{U_r}}}\left( n \right) + \frac{{\sum\limits_{k = 1}^K {{\theta _k}\left( n \right){P_k}\left( n \right)} {\rho _0}}}{{{d_{S{U_r}}} + H_1^2}}} \right\}}  \le {\Gamma _r},\forall r, 	\label{eq:p1.3c}\\
			&{d_{SE}} \le {\left( {\left\| {{{\mathbf{q}}_S}\left( n \right) - {{{\mathbf{\hat w}}}_E}} \right\| - {r_E}} \right)^2},	\label{eq:p1.3d}\\
			&{d_{S{U_r}}} \le {\left\| {{{\mathbf{q}}_S}\left( n \right) - {{\bf{w}}_{{U_r}}}} \right\|^2}, \forall r, 	\label{eq:p1.3e}\\
			&(\textrm{\ref{eq:C7}}), (\textrm{\ref{eq:C8}}),	\label{eq:p1.3f}
	\end{align}	
\end{subequations}
where
\begin{equation}
			{\hat R_{S{D_k}}}\left( n \right) \ge {L^m}\left( n \right)\left( {{{\left\| {{{\mathbf{q}}_S}\left( n \right) - {{\bf{w}}_{{D_k}}}} \right\|}^2} - {{\left\| {{\mathbf{q}}_S^m\left( n \right) - {{\bf{w}}_{{D_k}}}} \right\|}^2}} \right) + {D^m}\left( n \right),\\
\end{equation}
\begin{equation}
	{D^m}\left( n \right) = {\log _2}\left( {1 + \frac{{c\left( n \right){\rho _0}}}{{{{\left\| {{\mathbf{q}}_S^m\left( n \right) - {{\bf{w}}_{{D_k}}}} \right\|}^2} + H_1^2}}} \right),		\label{Dm} 	\\
\end{equation}
\begin{equation}
	{L^m}\left( n \right) = - \frac{{c\left( n \right){\rho _0}}}{{\ln (2)\left( {{{\left\| {{\mathbf{q}}_{S}^m\left( n \right) - {{\bf{w}}_{{D_k}}}} \right\|}^2} + H_1^2 + c\left( n \right){\rho _0}} \right)\left( {{{\left\| {{\mathbf{q}}_{S}^m\left( n \right) - {{\bf{w}}_{{D_k}}}} \right\|}^2} + H_1^2} \right)}}.	\label{Lm}
\end{equation}

It must be noted that (\ref{eq:p1.3d}) and (\ref{eq:p1.3e}) are non-convex functions since right-hand-side of them are convex function. By applying SCA, the following convex constraints are utilized
\begin{equation}
	\begin{aligned}
		{d_{SE}}  &\le 2{\left( {{\mathbf{q}}_S^m\left( n \right) - {{{\mathbf{\hat w}}}_E}} \right)^T}\left( {{{\mathbf{q}}_S}\left( n \right) - {\mathbf{q}}_S^m\left( n \right)} \right) \\
		&\;\;\;\;\; {\left\| {{\mathbf{q}}_S^m\left( n \right) - {{{\mathbf{\hat w}}}_E}} \right\|^2} - 2{r_E}\left\| {{\mathbf{q}}_S^m\left( n \right) - {{{\mathbf{\hat w}}}_E}} \right\| + {r_E}^2,
		\label{ddSE}
	\end{aligned}
\end{equation}
\begin{equation}
	{d_{S{U_r}}} \le {\left\| {{\mathbf{q}}_S^m\left( n \right) - {{\bf{w}}_{{U_r}}}} \right\|^2} + 2{\left( {{\mathbf{q}}_S^m\left( n \right) - {{\bf{w}}_{{U_r}}}} \right)^T}\left( {{{\mathbf{q}}_S}\left( n \right) - {\mathbf{q}}_S^m\left( n \right)} \right).
	\label{dSUr}
\end{equation}
By replacing (\ref{eq:p1.3d}) and (\ref{eq:p1.3e}) with (\ref{ddSE}) and (\ref{dSUr}), respectively, $\mathcal{P}_{1.3}$ is reformulated as
\begin{equation}
	\begin{aligned}
		\mathcal{P}_{1.3} \,:\, \max\limits_{{{\mathbf{q}}_S},{\eta _k} ,{d_{SE}},{{d_{S{U_r}}}}}\; & \sum\limits_{k = 1}^K {\eta _k}  \\
				{\mathrm{s.t.}}\; & (\textrm{\ref{eq:p1.3b}}),(\textrm{\ref{eq:p1.3g}}),(\textrm{\ref{eq:p1.3c}} ),( \textrm{\ref{ddSE}} ),( \textrm{\ref{dSUr}} ),( \textrm{\ref{eq:p1.3f}} ).
	\end{aligned}
\end{equation}
Now $\mathcal{P}_{1.3} $ is a convex problem and can be solved by existing optimization tools such as CVX.
\subsection{Subproblem 1.4: Optimizing Trajectory of $J$ }
Finally, we optimize ${{{\mathbf{q}}_J}\left( n \right)}$ with other variables fixed. Similarly, $\mathcal{P}_{1}$ is approximated as
\begin{subequations}
	\begin{align}		
		\mathcal{P}_{1.4} \,:\, \max\limits_{{{\mathbf{q}}_J},{\eta _k},{d_{J{U_r}}},{d_{JE}} }\; &\sum\limits_{k = 1}^K {{{\eta _k}}} \\
		{\mathrm{s.t.}}\; &\frac{1}{N}\sum\limits_{n = 1}^N {{\theta _k}\left( n \right){R_{\sec,2}^k}\left( n \right)}  \ge {{\eta _k}},\label{eq:p1.4b}\\
		&{\eta _k} \ge {R_{\min }}, \label{eq:p1.4f}\\
		&{{\mathbf{q}}_J}\left( 1 \right) = {\mathbf{q}}_J^0,{{\mathbf{q}}_J}\left( n \right) = {\mathbf{q}}_J^F,\label{eq:p1.4c}\\
		&\frac{1}{N}\sum\limits_{n = 1}^N {\left\{ {\frac{{{P_J}\left( n \right){\rho _0}}}{{{d_{J{U_r}}} + H_2^2}} + \sum\limits_{k = 1}^K {{\theta _k}\left( n \right){P_k}\left( n \right){h_S}_{{U_r}}\left( n \right)} } \right\}}  \le {\Gamma _r},\forall r,\label{eq:p1.4d}\\
		&{d_{J{U_r}}} \le {\left\| {{{\mathbf{q}}_J}\left( n \right) - {{\bf{w}}_{{U_r}}}} \right\|^2},	\label{eq:p1.4e}
	\end{align}
	\label{eq:P1.4}
\end{subequations}
where
\begin{equation}
	\begin{aligned}
		{R_{\sec,2}^k}\left( n \right) 
		&= {\log _2}\left( {1 + \frac{{{P_k}\left( n \right){h_{S{D_k}}}}}{{{\sigma ^2}}}} \right) - {\log _2}\left( {\frac{{{P_J}\left( n \right){\rho _0}}}{{{d_{JE}} + H_2^2}} + {\sigma ^2} + {P_k}\left( n \right){h_{SE}}} \right)\\
		&+ {\log _2}\left( {\frac{{{P_J}\left( n \right){\rho _0}}}{{{{\left( {\left\| {{{\mathbf{q}}_J}\left( n \right) - {{{\mathbf{\hat w}}}_{\mathrm{E}}}} \right\| + {r_E}} \right)}^2} + H_2^2}} + {\sigma ^2}} \right),
		\label{RSEC,2}
	\end{aligned}
\end{equation}
where ${{d_{JE}}}$ is a new auxiliary variable and satisfies the following constraints
\begin{equation}
 {d_{JE}} \le {\left( {\left\| {{{\mathbf{q}}_J}\left( n \right) - {{{\mathbf{\hat w}}}_E}} \right\| + {r_E}} \right)^2}.
 \label{dJE}
\end{equation}
By SCA, non-concave form in (\ref{RSEC,2}) is converted as
\begin{equation}
\begin{aligned}
	{R}_{\sec,2}^{\textrm{lb}}\left( n \right) 
	&= {\log _2}\left( {1 + \frac{{{P_k}\left( n \right){h_{S{D_k}}}}}{{{\sigma ^2}}}} \right) - {\log _2}\left( {\frac{{{P_J}\left( n \right){\rho _0}}}{{{d_{JE}} + H_2^2}} + {\sigma ^2} + {P_k}\left( n \right){h_{SE}}} \right)\\
	&+ N^m\left( n \right) + M^m\left( n \right)\left( {{{\left( {\left\| {{{\mathbf{q}}_J}\left( n \right) - {{{\mathbf{\hat w}}}_E}} \right\| + {r_E}} \right)}^2} - {{\left( {\left\| {{\mathbf{q}}_J^m\left( n \right) - {{{\mathbf{\hat w}}}_E}} \right\| + {r_E}} \right)}^2}} \right),
\label{R^{lb}}
\end{aligned}
\end{equation}
where
\begin{equation}	
{M^m}\left( n \right) = \frac{{{{\left( {\frac{{{P_J}\left( n \right){\rho _0}}}{{{{\left( {\left\| {{\bf{q}}_J^m\left( n \right) - {{{\bf{\hat w}}}_E}} \right\| + {r_E}} \right)}^2} + H_2^2}} + {\sigma ^2}} \right)}^{ - 1}}}}{{\ln (2)}}
\label{M_1^m},
\end{equation}
and
\begin{equation}
N^m\left( n \right) = {\log _2}\left( {\frac{{{P_J}\left( n \right){\rho _0}}}{{{{\left( {\left\| {{\mathbf{q}}_J^m\left( n \right) - {{{\mathbf{\hat w}}}_E}} \right\| + {r_E}} \right)}^2} + H_2^2}} + {\sigma ^2}} \right).
\label{N_1^m}
\end{equation}
Similar to $\mathcal{P}_{1.3}$, (\ref{eq:p1.4e}) and (\ref{dJE}) are approximated as
\begin{equation}
	\begin{aligned}
	r_E^2 \le {d_{JE}} &\le {\left\| {{\bf{q}}_J^m\left( n \right) - {{{\bf{\hat w}}}_E}} \right\|^2} + r_E^2 \\
	&+ \left( {2{{\left( {{\bf{q}}_J^m\left( n \right) - {{{\bf{\hat w}}}_E}} \right)}^T} + \frac{{2{r_E}{{\left( {{\bf{q}}_J^m\left( n \right) - {{{\bf{\hat w}}}_E}} \right)}^T}}}{{\left\| {{\bf{q}}_J^m\left( n \right) - {{{\bf{\hat w}}}_E}} \right\|}}} \right)\left( {{{\bf{q}}_J}\left( n \right) - {\bf{q}}_J^m\left( n \right)} \right),
		\label{dJE2}
	\end{aligned}
\end{equation}
and
\begin{equation}
	{d_{J{U_r}}} \le {\left\| {{{\mathbf{q}}_J}^m\left( n \right) - {{\bf{w}}_{{U_r}}}} \right\|^2} + 2{\left( {{{\mathbf{q}}_J}^m\left( n \right) - {{\bf{w}}_{{U_r}}}} \right)^T}\left( {{{\mathbf{q}}_J}\left( n \right) - {{\mathbf{q}}_J}^m\left( n \right)} \right),
	\label{dJUr}
\end{equation}
respectively.
Then $\mathcal{P}_{1}$ is reformulated as
\begin{equation}
\begin{aligned}
	\mathcal{P}_{1.4} \,:\, \max\limits_{{{\mathbf{q}}_J},{\eta _k},{d_{J{U_r}}}, {d_{JE}}} \;& \sum\limits_{k = 1}^K {{{\eta _k}}} \\
	{\mathrm{s.t.}}\; & (\textrm{\ref{eq:p1.4b}} ),(\textrm{\ref{eq:p1.4f}} ),(\textrm{\ref{eq:p1.4c}} ),( \textrm{\ref{eq:p1.4d}} ),(\textrm{\ref{dJE2}} ), \mathrm{and} \; (\textrm{\ref{dJUr}} ).
\end{aligned}
\end{equation}
$\mathcal{P}_{1.4}$ is a convex problem and can be solved by existing optimization tools such as CVX.

\begin{algorithm}[tb]
	\caption{Iterative Algorithm for Problem ($\mathcal{P}_{1}$)}
	\KwIn{Initialize feasible points}
	\While
	{$R\left( {{{\mathbf{\Theta }}^m},{{\mathbf{P}}^m},{\mathbf{q}}_S^m,{\mathbf{q}}_J^m} \right) - R\left( {{{\mathbf{\Theta }}^{m - 1}},{{\mathbf{P}}^{m - 1}},{\mathbf{q}}_S^{m - 1},{\mathbf{q}}_J^{m - 1}} \right)  \succ  {\varepsilon }$}
	{1. Solve ($\mathcal{P}_{1.1}$) for given $\left\{ {{{\mathbf{P}}^m},{\mathbf{q}}_S^m,{\mathbf{q}}_J^m} \right\}$ and obtain the solution ${{{\mathbf{\Theta }}^{m + 1}}}$;\\
	2. Solve ($\mathcal{P}_{1.2}$) for given $\left\{ {{{\mathbf{\Theta }}^{m + 1}},{\mathbf{q}}_S^m,{\mathbf{q}}_J^m} \right\}$ and obtain the solution ${{{\mathbf{P}}^{m + 1}}}$;\\
	3. Solve ($\mathcal{P}_{1.3}$) for given $\left\{ {{{\mathbf{\Theta }}^{m + 1}},{{\mathbf{P}}^{m + 1}},{\mathbf{q}}_J^m} \right\}$ and obtain the solution ${{\mathbf{q}}_S^{m + 1}}$;\\
	4. Solve ($\mathcal{P}_{1.4}$) for given $\left\{ {{{\mathbf{\Theta }}^{m + 1}},{{\mathbf{P}}^{m + 1}},{\mathbf{q}}_S^{m + 1}} \right\}$ and obtain the solution ${{\mathbf{q}}_J^{m + 1}}$;\\
	4. $m = m + 1$;\\
	5. Compute the objective value $R\left( {{{\mathbf{\Theta }}^m},{{\mathbf{P}}^m},{\mathbf{q}}_S^m,{\mathbf{q}}_J^m} \right)$.\\
	}
\KwOut{$R\left( {{{\mathbf{\Theta }}^m},{{\mathbf{P}}^m},{\mathbf{q}}_S^m,{\mathbf{q}}_J^m} \right)$ with ${{\mathbf{\Theta }}^*} = {{\mathbf{\Theta }}^{m}},\;{{\mathbf{P}}^*} = {{\mathbf{P}}^m},\;{\mathbf{q}}_S^* = {\mathbf{q}}_S^m,{\mathbf{q}}_J^* = {\mathbf{q}}_J^m$. }
\end{algorithm}

\subsection{Overall Algorithm, Convergence, and Complexity}
An iterative algorithm is proposed to solve $\mathcal{P}_{1}$ through combining four subproblems discussed above and applying BCD.
We need to select the initial feasible points and obtain the suboptimal solution by solving $\mathcal{P}_{1.1}$, $\mathcal{P}_{1.2}$, $\mathcal{P}_{1.3}$, and  $\mathcal{P}_{1.4}$ alternatively.  The obtained solutions in each iteration are utilized as the feasible input points for the next iteration.
Defining the objective function of the original problem $\mathcal{P}_{1}$ at the $m$th iteration as $R\left( {{{\mathbf{\Theta }}^m},{{\mathbf{P}}^m},{\mathbf{q}}_S^m,{\mathbf{q}}_J^m} \right)$,
\textbf{Algorithm 1} summarizes  the details of overall iterations for $\mathcal{P}_{1}$, 
where $\varepsilon$ denotes the tolerance of convergence.
The convergence of \textbf{Algorithm 1} is proved as follows.

\begin{proof}

In Step 1 of \textbf{Algorithm 1}, a standard linear problem $\mathcal{P}_{1.1}$ is solved and we obtain the solution ${{\mathbf{\Theta }}^{m + 1}}$. Thus, we have
\begin{equation}
	R\left( {{{\mathbf{\Theta }}^m},{{\mathbf{P}}^m},{\mathbf{q}}_S^m,{\mathbf{q}}_J^m} \right) \le R\left( {{{\mathbf{\Theta }}^{m+1}},{{\mathbf{P}}^m},{\mathbf{q}}_S^m,{\mathbf{q}}_J^m} \right).
\end{equation}
Then, the suboptimal solution ${{{\mathbf{P}}^{m + 1}}}$ is obtained through solving $\mathcal{P}_{1.3}$ as
\begin{equation}
	\begin{aligned}
		R\left( {{{\mathbf{\Theta }}^m},{{\mathbf{P}}^m},{\mathbf{q}}_S^m,{\mathbf{q}}_J^m} \right)&= R\left( {{{\mathbf{\Theta }}^{m+1}},{{\mathbf{P}}^m},{\mathbf{q}}_S^m,{\mathbf{q}}_J^m} \right)\\
		&  \le {R^{lb}}\left( {{{\mathbf{\Theta }}^{m+1}},{{\mathbf{P}}^{m+1}},{\mathbf{q}}_S^m,{\mathbf{q}}_J^m}\right)\\
		&  \le R\left( {{{\mathbf{\Theta }}^{m+1}},{{\mathbf{P}}^{m+1}},{\mathbf{q}}_S^m,{\mathbf{q}}_J^m}\right),
	\end{aligned}
	\label{63}
\end{equation}
where ${R^{lb}}\left( {{{\mathbf{\Theta }}^{m+1}},{{\mathbf{P}}^{m+1}},{\mathbf{q}}_S^m,{\mathbf{q}}_J^m}\right)$ is the objective function of the approximate problem $\mathcal{P}_{1.3}$, which is a lower bound to the objective function of $\mathcal{P}_{1}$. Eq. (\ref{63}) indicates that the objective function is always non-decreasing after each iteration.

The proof of the convergence in Step 3 and Step 4 is similar to that of (\ref{63}), and the result follows
\begin{equation}
	R\left({{{\mathbf{\Theta }}^m},{{\mathbf{P}}^m},{\mathbf{q}}_S^m,{\mathbf{q}}_J^m} \right) \le R\left( {{{\mathbf{\Theta }}^{m + 1}},{{\mathbf{P}}^{m + 1}},{\mathbf{q}}_S^{m + 1},{\mathbf{q}}_J^m} \right),
\end{equation}
and
\begin{equation}
	R\left({{{\mathbf{\Theta }}^m},{{\mathbf{P}}^m},{\mathbf{q}}_S^m,{\mathbf{q}}_J^m} \right) \le R\left( {{{\mathbf{\Theta }}^{m + 1}},{{\mathbf{P}}^{m + 1}},{\mathbf{q}}_S^{m + 1},{\mathbf{q}}_J^{m + 1}} \right).
\end{equation}
Thus, we further find the objective function in $\mathcal{P}_{1}$ is non-decreasing after each iteration, which is upper bounded by a finite value due to the feasible set under the constraints. Then \textbf{Algorithm 1} is convergent.
\end{proof}
The complexity of \textbf{Algorithm 1} comes from two
aspects. The first aspect is about the complexity analysis of the user scheduling coefficient, which is solved via standard linear programming by the interior point method with complexity ${\cal O}\left( {\sqrt {N\left( {K + 1} \right)} \log \frac{1}{\varepsilon }} \right)$.
The second aspect is about the complexity analysis of solving  transmit power and UAV trajectory optimization by using SCA, where the complexity is ${\cal O}\left( {{{\left( {\left( {K + R + 1} \right)N} \right)}^{3.5}}\log \frac{1}{\varepsilon }} \right)$.
It is assumed that the number of iterations in the outer and inner loops are $M_1$ and ${M_2}$. Thus, the total
complexity of \textbf{Algorithm 1} is
${\cal O}\left( {{M_1}\sqrt {N\left( {K + 1} \right)} \log \frac{1}{\varepsilon } + {M_2}{{\left( {\left( {K + R + 1} \right)N} \right)}^{3.5}}\log \frac{1}{\varepsilon }} \right)$.

\section{Numerical Results and Analysis}
\label{sec:Simulation}

In this section, numerical results are provided for verifying the convergence and effectiveness of the proposed algorithm.
We consider a 2-D square area  with three CUs ($K = 3$), two PUs ($R = 2$), and an Eve.
The horizonal coordinates are set as $ {\left[ { - 55, - 10}; { 0, - 65}; { 50, - 5} \right]^T}$ and ${\left[ {30,25}; {-30,25} \right]^T}$.
The center of the circle of the area where $E$ is located is set as ${{\bf{\hat w}}_E} = {\left[ {15, - 15} \right]^T}$.
According to \cite{WangY2021TCCN}, the maximum flight speed of UAVs is set as  ${V_{\max }} = 7$ m/s and the altitudes of $S$ and $J$ are set as ${H_1} = 15$ m and ${H_2} = 10$ m, respectively.
The channel power gain at a reference distance of 1 m is ${\rho _0} =  - 30$ dBm and the noise power is set as ${\sigma ^2} =  - 90$ dBm. The maximum transmit power levels of the UAVs are $P_S^{\max } = 4P_S^{ave}$ and $P_J^{\max } = 4P_J^{ave}$, respectively. The average interference power threshold at the PU is set as ${\Gamma _r} =  - 80$ dBm and and the error tolerance is set as ${\varepsilon} = 0.001$.

The numerical results of three benchmark schemes are presented in the simulations.
\begin{enumerate}

\item Benchmark I: Both $S$ and $J$ utilize constant transmit power and follow a circular flight path.

\item Benchmark II: $S$ is designed with joint optimization of power and trajectory, $J$ keeps flying with a constant power and circular trajectory.

\item Benchmark III: $J$ takes a joint optimization of power and trajectory to disrupt  $E$, and $S$ flies with constant transmitting power and circular flight trajectory.

\end{enumerate}

The initial flight trajectory of UAVs is set as circular with constant speed whose center is the geometric center of ground nodes, i.e., $C = \frac{1}{K}\sum\limits_{k = 1}^K {{{\bf{w}}_{{D_k}}}} $. 
It is suitable for ${S}$ to visit all CUs and its initial flight radius for $S$ can be computed according to ${{{R}}_S} = \min \left( {\frac{{{{{V}}_{\max }}T}}{{2\pi }},\left\| {C - {{\bf{w}}_{{D_k}}}} \right\|} \right)$.
The initial flight path of the ${J}$ must cover region where the eavesdropping node is uncertain. 
Therefore, its initial flight radius for $S$ is defined as ${R_J} = \frac{1}{2}{R_S}$. Based on initial flight trajectories, the initial power is set as $ P_S^0 =  P_S^{ave}$ and $ P_J^0 =  P_J^{ave}$.
Similar to \cite{ZhangG2019TWC}, if ${{R_{S{D_k}}}\left( n \right) - {R_E}\left( n \right)}$ is non-negative, $ P_S = 0$.

\begin{figure}[t]
	\centering
	\includegraphics[width = 3.5 in]{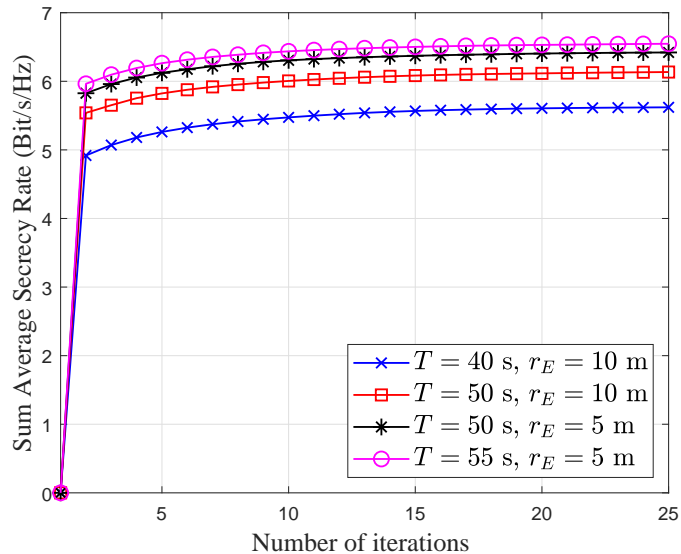}
	\caption{Relationship between the sum average secrecy rate and the number of iterations.}
	\label{fig2iterative}
\end{figure}
Fig. \ref{fig2iterative} demonstrates the sum average secrecy rate of the considered system versus the number of iterations for UAVs in different flight periods under varying uncertain eavesdropping coverage. The results demonstrate the convergence of the proposed scheme regarding the jointly optimized user scheduling coefficients, the UAV transmitting power, and the UAV flight trajectory. One can observe that the average sum secrecy rate increases quickly with the flight time, the number of iterations, and converges within around 15 iterations in the listed four flight periods. Furthermore, as the range of eavesdropping uncertainty area decreases, the system's average sum secrecy rate gradually increases.

\begin{figure}[t]
	\centering
	\includegraphics[width = 3.5 in]{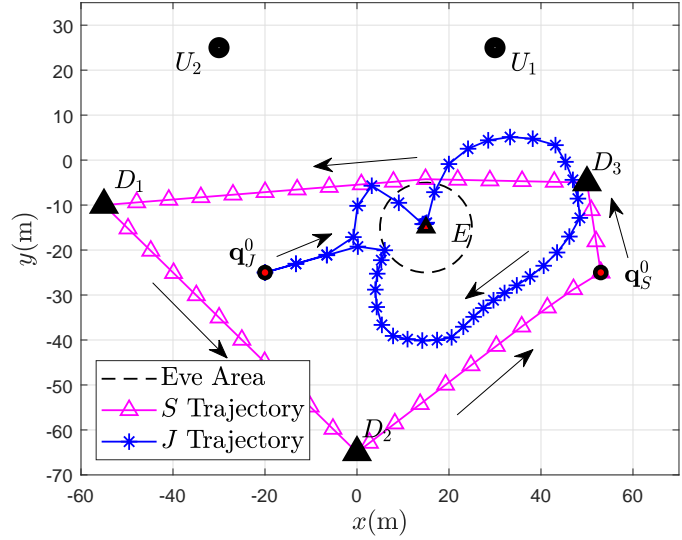}
	\caption{The optimal trajectory of UAVs during the flight cycle.}
	\label{fig3trajectory}
\end{figure}
Fig. \ref{fig3trajectory} demonstrates the optimal UAV trajectories achieved by jointly optimizing user scheduling, UAV transmitting power, and flight trajectories, in which the black arrows denote the flight directions of the two UAVs, which are in opposite directions.
By adopting the proposed iterative algorithm, the flight trajectory of UAVs eventually changes from the initial circular trajectory to the flight trajectory.
It can be observed that $S$ starts from the departure point and gradually approaches, hovers, and leaves each cognitive user, which means the approach of scheduling CUs within each time slot is utilized by the base station in the considered system using the frequency band authorized by the primary network. Moreover, based on the trajectory of the jammer UAV, although the location of $E$ cannot be obtained precisely, an optimal trajectory is obtained to fly around the area where $E$ is located through a joint optimization algorithm to better suppress eavesdropping the legitimate signals.
\begin{figure}[t]
	\centering
	\includegraphics[width = 3.5 in]{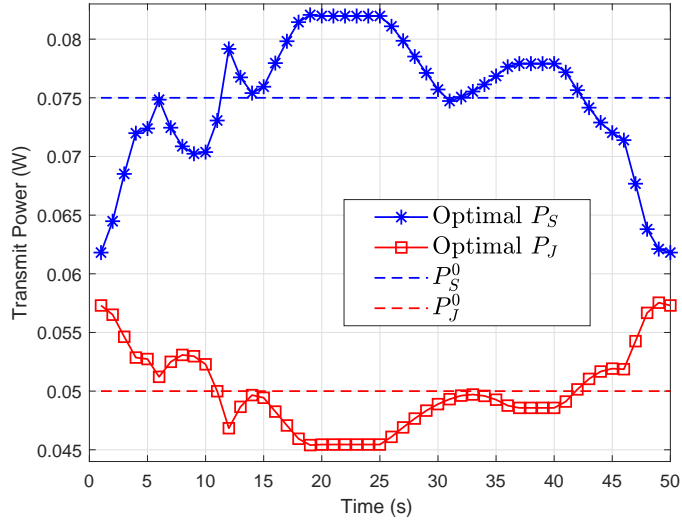}
	\caption{Dynamic power variation during the flight cycle of the UAVs.}
	\label{fig8power}
\end{figure}
Fig. \ref{fig8power} illustrates the dynamic variation of the UAV transmitting power during a flight cycle.
One can observe ${P_S}$ and ${P_J}$ keep a tendency to couple. When ${P_J}$ experiences an upward trend, ${P_S}$ experiences a downward trend.

\begin{figure}[t]
	\centering
	\includegraphics[width = 3.5 in]{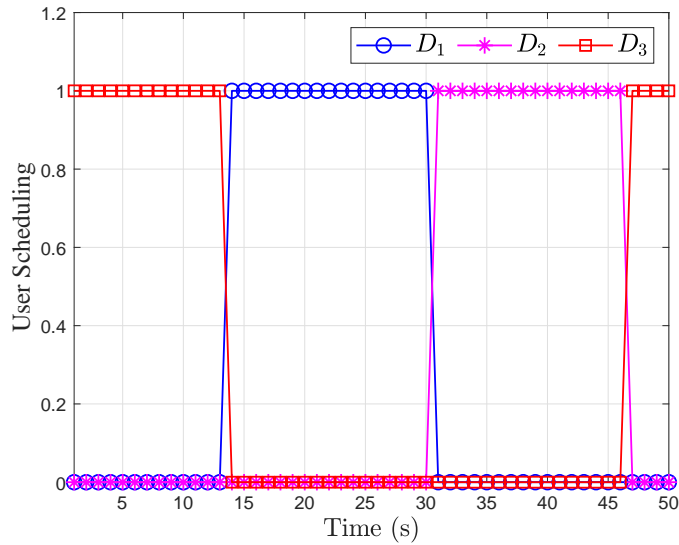}
	\caption{User scheduling during the UAVs flight cycle.}
	\label{fig4Scheduling}
\end{figure}
The user scheduling result is plotted in Fig. \ref{fig4Scheduling}. It can be observed that all the CUs are alternately awake, which indicates that only one user is scheduled by $S$ in each time slot, at which point the unscheduled users keep silent.
The UAV hovers over the target area for a period, all the CUs obtain service in turn. Since UAV cannot completely cover users in a time slot, it is necessary to optimize the user scheduling to ensure the secure communication on each user. Therefore, the minimum average security rate must be satisfied as (\ref{eq:C10}).
For all users satisfying secure communication at the same slot, the user with the largest secure rate will be scheduled.
Since the average secrecy rate is optimized considering the minimum secrecy rate on all the CUs, fairness is achieved among these CUs, as shown in Fig. \ref{fig4Scheduling}.

\begin{figure}[t]
	\centering
	\includegraphics[width = 3.5 in]{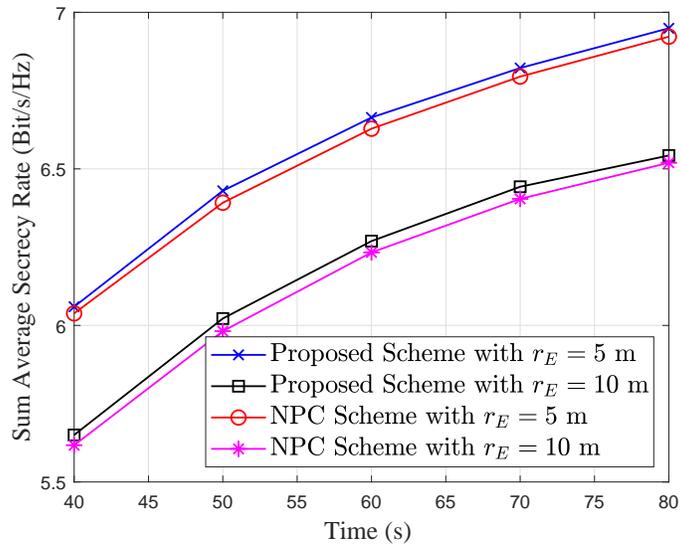}
	\caption{Performance comparison of the proposed scheme and the NPC scheme with varying ${{r_E}}$.}
	\label{fig5Comparison}
\end{figure}
In Fig. \ref{fig5Comparison}, we compare the performance between the proposed scheme and no power control (NPC) scheme with varying time and radius of the eavesdropper uncertainty area. One can observe the average secrecy rate has an increasing trend with increasing flight time of the UAVs and/or performing power control. The eavesdropper uncertainty area radius, which reflects the CSI between the UAV and the eavesdropping node, has a considerable impact on the average security rate of the considered system. The smaller the ${r_E}$, the clearer the base station is about the location information of the eavesdropper.

\begin{figure}[t]
	\centering
	\includegraphics[width = 3.5 in]{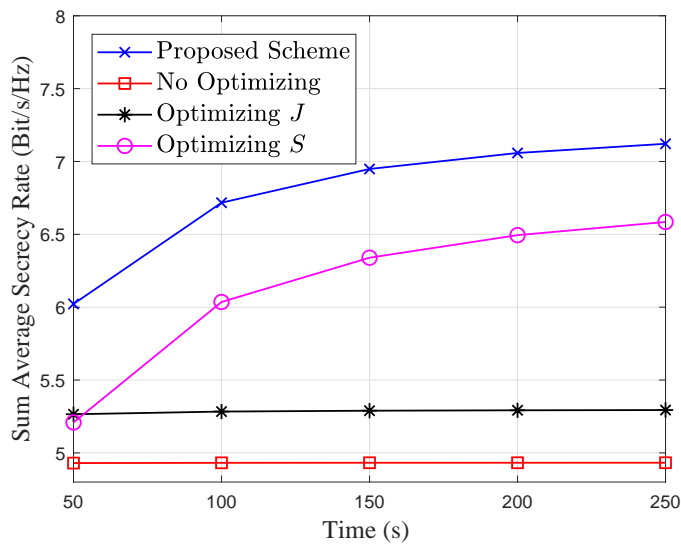}
	\caption{The sum average secrecy rate versus $T$ with different scenarios.}
	\label{fig6benchmark}
\end{figure}
Fig. \ref{fig6benchmark} demonstrates the variation of the secrecy rate of the system with time $T$ for different scenarios wherein constant transmit power is utilized on $S$ and/or $J$.
%
%
%
One can clearly observe that the sum secrecy rate of the system obtained by the proposed iterative algorithm based on BCD and SCA is much better than the other scenarios.
Comparing with optimizing trajectory of $S$ alone, optimizing both the power and trajectory of $J$ can effectively cope with the uncertainty of the eavesdropper location. The scheme proposed in this work, i.e., optimizing the power and trajectory of both $S$ and $J$ can maximize the security performance of the considered system.

%

\begin{figure}[t]
	\centering
	\includegraphics[width = 3.5 in]{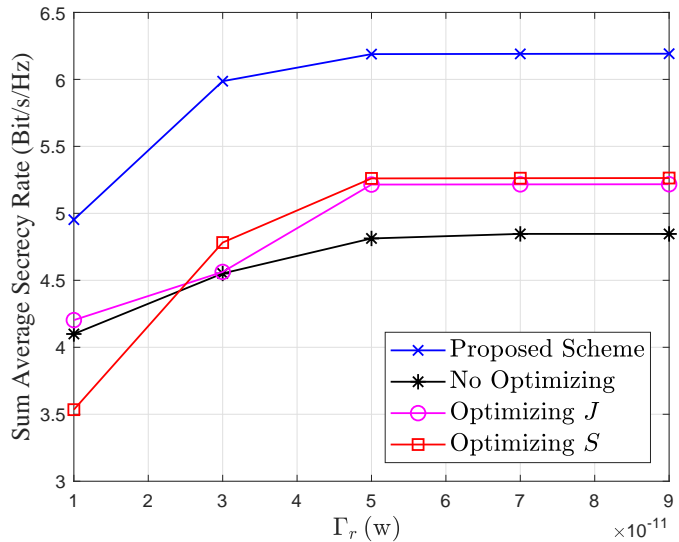}
	\caption{The sum average secrecy rate versus interference threshold constraint ${\Gamma _r}$ with different scenarios.}
	\label{fig7underlay}
\end{figure}
Fig. \ref{fig7underlay} illustrates the effect of the interference power thresholds of the PUs on the average secrecy rate concerning the flight time under different scenarios.
It is observed that the average secrecy rate achieved by all schemes increases rapidly in the lower-${\Gamma _r}$ region then tends to be saturated. This is because the available transmit power in the lower-${\Gamma _r}$ region increases with the relaxation of the average interference power thresholds constraints, then $S$ has a higher degree of freedom in power allocation. Thus, the secrecy rate is effectively improved. However, in the larger-${\Gamma _r}$ region, the transmit power on $S$ and $J$ are restricted by $P_S^{\max }$ and  $P_J^{\max }$ to meet the maximum power constraints rather than interference threshold constraints.

\section{Conclusion}
\label{sec:Conclusions}
This paper investigates a dual collaborative UAV system for physical layer security in a multi-user scheduling scenario
on the uncertainty of potential eavesdropping node locations in CRNs. The joint design of maximizing the average secrecy rate is formulated as a mixed-integer non-convex problem by considering UAV trajectories, user scheduling, and transmit power. An efficient iterative algorithm based on BCD and SCA was proposed to tackle the challenging non-convex problem and achieve a suboptimal solution. Numerical results verified the convergence and effectiveness of our proposed algorithm.

\end{document}